SUPERSTITION: BELIEF IN THE AGE OF SCIENCE
by Robert Park, University of Maryland. Princeton University Press (2008)

**Reviewed by Adrian L. Melott**

*Adrian L. Melott is professor of physics and astronomy at the University of Kansas. He is the author of over 100 peer-reviewed journal articles in physical cosmology, but since 2003 has moved to "astrobiophysics", in which he examines possible astrophysical effects on the biosphere—much of it in collaboration with paleontologists. He is a Fellow of the APS and the AAAS.*

Bob Park was saved by a miracle. He's not at all young. It's God's will that he is alive at all. A big tree fell on him as he passed by, and after a long spell in the hospital he returned from the dead. God brought him back so he could write eloquently to tell us all what a bunch of crap this kind of thinking is.

I could end the review here, but I suppose many would find that somehow lacking, so I will go on. Bob is a distinguished physicist who devotes himself to writing, mostly about the nonsense he sees in the world. He has a brief weekly topical news-editorial email which can be accessed at http://www.bobpark.org/ which applies his acerbic wit to all kinds of things from perpetual motion cons to the space station to energy policy and population issues. The book I review here is written in much the same style, with much of the same kind of appeal.

The book shows no particular respect for its targets, one of which is religion. Thus targets of his attack include not only the fundamentalists who attack evolution, but also the Templeton Foundation who seek to find and promote commonality between science and religion, and the physicists who promote fine-tuning arguments and the anthropic principle and are financially rewarded for it with funding in excess of the Nobel Prize. Other targets include alternative medicine and closely-related New Age beliefs, quantum consciousness mysticism, recovered memories, the medical efficacy of intercessory prayer, the alleged religious base of morality, environmental problems, overreaching technological optimism, and more. However, he does show some respect for certain individuals with whom he disagrees, symbolized by a pair of Catholic priests named David and Shaun who reappear throughout the book.

The book is written in the same style as his weekly column, which I would describe as "deceptively simple". It has none of the literary elegance or complexity we associate with some of the best science writers, but it is of equally high quality. If I were to look for a model in fiction, it would be Kurt Vonnegut. Thus, the writing is broadly accessible without insulting one's intelligence, which is extremely valuable in this sort of book.

I have some disagreements with Park. Like Dawkins, Weinberg, and many others he equates religion with "believing things". He rightly notes the absurd and/or damaging beliefs associated with many of the world's religions.  This is a very Western-centric interpretation of religion.  There exist major religious groups for whom following some set of laws is what matters; there are others that emphasize meditational practices or simply love.  He is right that most religions devolve to cult-like or magical practice, but this is not universal.

In his discussion of alternative medical research, he emphasizes the avoidance of double-blind, placebo-controlled, statistically significant research.  While it is admittedly not superstition and therefore off-topic, he nevertheless fails to do more than note in passing the extent to which funding by the pharmaceutical industry corrupts research in mainstream medicine.  We all know the stories of side effects which were suppressed in early studies and then emerge to injure thousands of people.  This is a form of cherry-picking results, which he does discuss in the context of parapsychology, for example, so it would be fair game and would contribute to a balanced discussion.

The longest chapter deals with the attempts by some in the religious community to repress the teaching of evolution, told mostly from a US historical perspective.  It is condensed (as a single chapter must be) but tells the essentials from the days of Thomas Huxley, through the Scopes Trial, the evolution of creationism into "creation science", the mutation of intelligent design which allowed the movement to speciate and enter a new ecological niche in the US middle class, pioneer species such as Jonathan Wells, and the Dover Trial.  It brings out all the important points and can be highly recommended.  It is valuable in that it is put into a broader perspective.

There's something here to offend almost everyone.  My New Age friends who support the teaching of evolution will be upset by the attack on their herbs. If they can get past that, they will enjoy this book, learn from it, and most importantly allow it to clarify their thinking.  It worked for me.

Opinions are the author's and not necessarily shared by NCSE, but they should be.